\documentclass[doublecol]{epl2}

\usepackage{graphicx,color}
\usepackage{bm}
\usepackage{amsmath}

\title{On lines of constant polarisation in structured light beams}

\author{S.M. Barnett\inst{1} \and F.C. Speirits\inst{1,*} \and J.B. G\"{o}tte\inst{1}}
\shortauthor{S.M. Barnett \etal}

\institute{
  \inst{1} School of Physics and Astronomy, University of Glasgow, Glasgow G12 8QQ, UK\\
  \inst{*} E-mail: \texttt{fiona.speirits@glasgow.ac.uk} (corresponding author)
}

\abstract{We show that Skyrmion field lines, constructed from the local Stokes parameters, trace out lines of constant optical polarisation.}

\begin{document}

\maketitle

Structured light beams are characterised by an engineered spatial variation of
amplitude, phase and polarisation \cite{Nye,Roberta,Dennis,Forbes}.  Important
examples of these include beams carrying orbital angular momentum
\cite{Allen,AllenBook,Bekshaev,SonjaRev,Alison}, helicity lattices
\cite{CohenT,Dalibard,RobC,RobCa,Koen,Brasselet}, and the vector vortex beams
\cite{Zhan,Neal,Zhan02,Dorn}. Some of these beams, in particular those with
spatially varying polarisation, have been shown to exhibit Skyrmionic structure
\cite{Scarlett,Amy}.  Typically, these have a polarisation pattern in the
transverse plane that, at its centre, has one polarisation but at the outer
reaches of the plane has the orthogonal polarisation. In general, all possible
polarisations appear at some point in this transverse plane in a winding
pattern, and Skyrmions are characterised by a corresponding winding number, the
Skyrmion number~\cite{Scarlett,Amy,Pisanty,Denz1,Claire,Shen1,Shen2}. There
exist numerous variations on this theme
\cite{Pisanty,Denz1,Claire,Shen1,Shen2}. What has not yet been identified,
however, is the physical significance of the Skyrmion field itself: we rectify
that in this letter.

We present an unexpected property of Skyrmion field lines that has application
whether or not a structured light beam has a non-zero Skyrmion number, as long
as the polarisation pattern covers the whole Poincar\'{e} sphere continuously
\cite{Miguel}. Put simply, it is that Skyrmion field lines trace out contours
of constant polarisation. Moreover, all such lines of constant polarisation are
Skyrmion field lines. Several important properties of structured light beams
then follow from the mathematical properties of the Skyrmion field.  The
central theme of our paper is the application of these ideas to paraxial light
beams, but we conclude with a brief discussion of these ideas in other fields
of physics, including electron \cite{El-Kareh,Klemperer,Hawkes} and neutron
\cite{Rauch} optics and also gravitational waves \cite{Maggiore}.

Skyrmion field lines for paraxial light beams are defined in terms of the
normalised Stokes parameters, $S_1$, $S_2$ and $S_3$ \cite{BW}.  The $i^{th}$
component of the Skyrmion field is \cite{Scarlett,Amy}

\begin{equation}
\label{Eq1}
\Sigma_i = \frac{1}{2}\varepsilon_{ijk}\varepsilon_{pqr}S_p\frac{\partial S_q}{\partial x_j}\frac{\partial S_r}{\partial x_k} \, ,
\end{equation}
where $\varepsilon_{ijk}$ is the alternating or Levi-Civita symbol and we
employ the summation convention in which a summation is implied over repeated
indices.  The specific form of $\Sigma_i$ is crucial to an appreciation of the
link with lines of constant polarisation.  For this reason it is worth writing
explicitly one of the Cartesian components of $\bm{\Sigma}$

\begin{eqnarray}
\label{Eq2}
\Sigma_z &=& \frac{1}{2}\varepsilon_{pqr}S_p\left(\frac{\partial S_q}{\partial x}\frac{\partial S_r}{\partial y}
- \frac{\partial S_r}{\partial x}\frac{\partial S_q}{\partial y}\right)  \nonumber \\
&=& S_1\left(\frac{\partial S_2}{\partial x}\frac{\partial S_3}{\partial y}
- \frac{\partial S_3}{\partial x}\frac{\partial S_2}{\partial y}\right)  \nonumber \\
& &  + S_2\left(\frac{\partial S_3}{\partial x}\frac{\partial S_1}{\partial y}
- \frac{\partial S_1}{\partial x}\frac{\partial S_3}{\partial y}\right) \nonumber \\
& & + S_3\left(\frac{\partial S_1}{\partial x}\frac{\partial S_2}{\partial y}
- \frac{\partial S_2}{\partial x}\frac{\partial S_1}{\partial y}\right) \, .
\end{eqnarray}
Note that this $z$-component depends on the variation of the Stokes parameters,
and therefore of the polarisations, only in the $x$ - and $y$ - directions.
Each term, moreover, depends on all three Stokes parameters.

The Skyrmion number associated with our structured beam is readily obtained by
integration over the plane transverse to the direction of propagation.  If we
take this direction to define our $z$ - axis, then the Skyrmion number is
\begin{equation}
\label{Eq3}
n = \frac{1}{4\pi} \int \Sigma_z \, dx dy \, ,
\end{equation}
where the integral runs over the whole transverse plane.  This value is
typically an integer, although structures with non-integer Skyrmion number can
be constructed~\cite{Scarlett}.  Our present concerns do not involve Skyrmions
or the Skyrmion number explicitly, but rather focus on the Skyrmion field. This
exists wherever there is a continuously spatially varying polarisation apart
from a few special cases, such as where there is a polarisation variation only
in one direction.  We note that the Skyrmion field is a transverse field in
that
\begin{equation}
\label{Eq4}
\bm{\nabla} \cdot \bm{\Sigma} = 0 \, ,
\end{equation}
and therefore the integral over any closed surface is zero $\oint \bm{\Sigma}
\cdot d\, \mathbf{S} = 0$.  The only exception to this condition will occur if
there are lines along which the polarisation is undefined.

\begin{figure}
    \begin{center}
        \includegraphics[width=\linewidth]{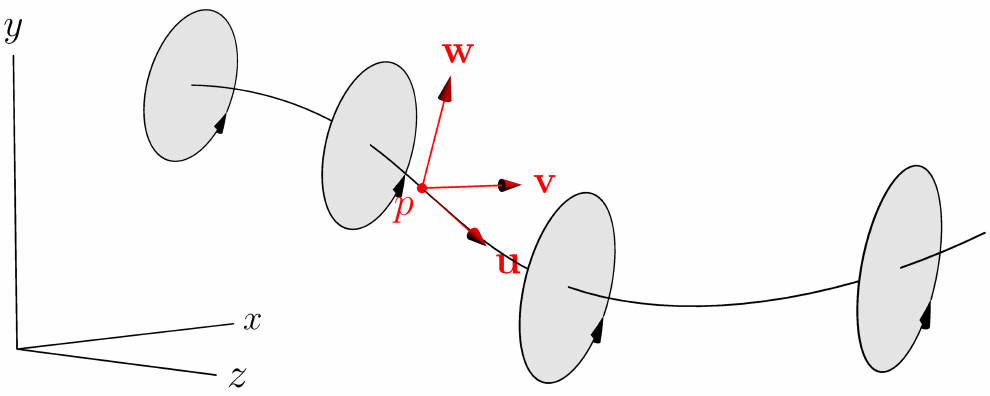}
    \end{center}
    \caption{\label{fig:fieldline} Plot of a line of constant elliptical
    polarisation and the local coordinate system $\mathbf{u}, \mathbf{v}, \mathbf{w}$ at $p$.}
\end{figure}

Let us turn to the properties of lines of constant polarisation.  As is well
known, structured light beams are threaded by lines of constant polarisation.
The most studied example is the C-lines along which the polarisation is purely
left- or right-handed circularly polarised \cite{Nye,Nye1983}.  There is
nothing in this context, however, that is specific to circular polarisation,
and we can trace such contours of constant polarisation for any chosen
polarisation.  Consider a point $p$ in a structured paraxial light beam, as
depicted in Fig. \ref{fig:fieldline}.  From this point there extends a line (in
two directions) along which the polarisation is the same as at $p$.  Note that
the amplitude and phase will not, in general, remain the same along this line.
Let us introduce a local right-handed Cartesian coordinate system
$(\mathbf{u,v,w})$ at $p$, in which the line of constant polarisation extends
in the direction $\mathbf{u}$. As the polarisation in the direction
$\mathbf{u}$ (and $-\mathbf{u}$) is unchanged, it follows that the direction of
the Stokes vector $\mathbf{S}$ is also unchanged:
\begin{equation}
\label{Eq5}
(\mathbf{u}\cdot \bm{\nabla}) \mathbf{S} = \frac{\partial}{\partial u}\mathbf{S} = 0 \, ,
\end{equation}
where ${\bf u}$ is a unit vector in the direction of the coordinate $u$. We can
write the components of the Skyrmion field at $p$ in the $u,v,w$ basis and find

\begin{eqnarray}
\label{Eq6}
\Sigma_u &=& \frac{1}{2}\varepsilon_{pqr}S_p\left(\frac{\partial S_q}{\partial v}\frac{\partial S_r}{\partial w}
- \frac{\partial S_r}{\partial v}\frac{\partial S_q}{\partial w}\right)  \nonumber \\
\Sigma_v &=& \frac{1}{2}\varepsilon_{pqr}S_p\left(\frac{\partial S_q}{\partial w}\frac{\partial S_r}{\partial u}
- \frac{\partial S_r}{\partial w}\frac{\partial S_q}{\partial u}\right) \\
\Sigma_w &=& \frac{1}{2}\varepsilon_{pqr}S_p\left(\frac{\partial S_q}{\partial u}\frac{\partial S_r}{\partial v}
- \frac{\partial S_r}{\partial u}\frac{\partial S_q}{\partial v}\right) \, . \nonumber
\end{eqnarray}
The derivatives of the Stokes parameters with respect to $u$ are zero and it
follows that $\Sigma_v = 0 = \Sigma_w$, and therefore that the Skyrmion field
line points in the direction of constant polarisation.  This is our principal
result.

It is straightforward to confirm that the Skyrmion field is independent of the
basis used to denote the Stokes vectors and hence the identification of the
Skyrmion field lines with lines of constant polarisation holds for every
possible polarisation.  Such a global transformation changes the polarisation
at every point in the field but does not alter the Skyrmion field, which is
associated with lines of constant polarisation, but not the specific
polarisation along these lines. Identifying a Skyrmion field line does not
determine the polarisation along the field line, merely the line itself. More
formally, the Skyrmion field is invariant under any unitary transformation of
the Poincar\'{e} sphere and so is not dependent on the basis used to express
$S_1$, $S_2$ and $S_3$. In this way, the Skyrmion field extends the
characterization of lines of constant circular or linear polarisation
\cite{BerryDennis} to every polarisation.


The mathematical properties of the Skyrmion field allow us to make general
statements about lines of constant polarisation.  The simplest and most
important among these follows from the transverse nature of the Skyrmion field,
$\bm{\nabla} \cdot \bm{\Sigma} = 0$. Like other transverse fields, such as the
magnetic induction $\mathbf{B}$ in electromagnetism, Skyrmion field lines
cannot start or end (no monopoles) and nor can they branch or coalesce.  The
identification of Skyrmion field lines with lines of constant polarisation
means that the same properties must hold for lines of constant polarisation.
The only exception to this rule occurs along lines at which the polarisation is
undefined, where several lines of different polarisation can meet.  At such
lines, however, the transverse condition on {\bm{$\Sigma$}} will fail.

One remaining subtlety needs to be addressed.  This is the fact that lines of
constant polarisation do not have a preferred sense of direction: such a line
is independent of whichever direction we choose to move along it.  The Skyrmion
field line, however, has a specific direction; if we change the sign of
{\bm{$\Sigma$}}, then it reverses its direction along the line of constant
polarisation.  In this sense, at least, there seems to be a significant
difference between lines of constant polarisation and Skyrmion field lines and
we should explain the origin of this difference.

We have seen that the Skyrmion field lines do not determine the local
polarisation, merely the local direction along which the polarisation does not
vary. For the structured light beam, however, there is a further class of
symmetries that leaves the pattern of lines of constant polarisation unchanged.
This is to apply the operation of complex conjugation to the polarisations.  To
be specific, let $\mathbf{e}_\text{H}$ and $\mathbf{e}_\text{V}$ be the real
unit vectors corresponding to horizontal and vertical polarisation, so that
left- and right-circular polarisations are $\mathbf{e}_\text{L} = (\mathbf
{e}_\text{H} + i \mathbf{e}_\text{V})/\sqrt{2}$ and $\mathbf{e}_\text{R} =
(\mathbf{e}_\text{H} - i \mathbf{e}_\text{V})/\sqrt{2}$.  If we apply the
complex conjugation operation then the left- and right-circular polarisations
switch, but the horizontal and vertical are unchanged, as are all the other
possible linear polarisations, and we arrive at an alternative (but physically
allowed) polarisation pattern with the same lines of constant polarisation.
This transformation is antiunitary in nature \cite{Wigner}.  Such
transformations are familiar from the study of time-reversal and CP symmetries
in particle physics \cite{Bigi}.  The complex conjugate transformation coupled
with rotations provides a second set of symmetries under which the polarisation
changes but the lines of constant polarisation do not.  The Skyrmion field
lines, however, switch direction under the antiunitary transformation as their
value is based on a right-handed coordinate system for the Poincar\'{e} sphere.
The complex conjugation operation applied to polarisation, however, changes a
right-handed arrangement of the Stokes parameters into a left-handed one and,
in doing so, flips the sign of the Skyrmion field.

It is interesting to ask if there is always a Skyrmion field in a structured
beam. The answer follows from the discussion in the preceding paragraph; if,
for example, the field is unchanged by performing complex conjugation on the
polarisation then the Skyrmion field is everywhere equal to its negative and
therefore is equal to zero. Simple examples are the radially and azimuthally
polarised beams \cite{Zhan} for which the polarisation is everywhere linear.
For such beams we have \emph{planes} of constant polarisation rather than
lines.

The connection between the Skyrmion field and lines of constant polarisation
has been established here for paraxial structured light beams, but we may
expect it to have wider applications.  For electrons and neutrons, a similar
association will hold for the Skyrmion field lines and lines along which the
particle spin does not change.  For non-paraxial optical fields there exists a
variety of features that can be associated with Skyrmions and, by extension,
with a Skyrmion field.  It will be interesting to see how these are related to
the spatial arrangement of spin-related properties of the electromagnetic
field.  Finally, gravitational waves have two orthogonal polarisations and so
we would expect Skyrmion field lines to be associated, also, with spatial
variations of the polarisation of these fields.

In summary, we have shown that lines of constant polarisation in any structured
paraxial light beam are identified with Skyrmion field lines.  It follows that
there is a more intimate relationship between structured light beams and
Skyrmion fields than simply whether or not a particular beam has an associated
Skyrmion number or Skyrmionic structures.

\acknowledgments This work was supported by a Royal Society Research
Professorship, grant number RP150122, and the UK Engineering and Physical
Sciences Research Council, grant numbers EP/R008264/1 and EP/V048449/1.
\newline

\emph{Data availability statement}: No new data were created or
analysed in this study.

\end{document}